\documentstyle[prl,aps,mypsfig2,twocolumn,floats]{revtex}

\begin{document}

\preprint{}


\title{Efficient implementation of selective recoupling in 
	heteronuclear spin systems using Hadamard matrices}

\author{Debbie W. Leung$^{1,2}$, 
 	Isaac L. Chuang$^2$, 
	Fumiko Yamaguchi$^{1}$ and 
	Yoshihisa Yamamoto$^{1,3}$} 

\address{\vspace*{1.2ex}
	{$^1$ ICORP Quantum Entanglement Project, 
		\\ Edward L. Ginzton Laboratory, Stanford University  
		\\ Stanford, CA 94305-4085}	\\[1.2ex]
 	{$^2$ IBM Almaden Research Center \\
		San Jose, CA 94120}	\\[1.2ex]
 	{$^3$ NTT Basic Research Laboratories 
		\\ 3-1 Morinosato-Wakamiya, Atsugi, Kanagawa 243-0198, Japan}
		\\[1.2ex]
}
	
\date{\today}
\maketitle

\def\<{\langle}
\def\>{\rangle}
\def\be{\begin{equation}}
\def\ee{\end{equation}}
\def\bea{\begin{eqnarray}}
\def\eea{\end{eqnarray}}
\def\lbL{ \left[\rule{0pt}{2.4ex} }
\def\rbL{ \right] }
\def\lbm{ \left[\rule{0pt}{2.1ex}\right. }
\def\rbm{ \left.\rule{0pt}{2.1ex}\right] }
\def\hs{ \hspace*{0.5ex}}

\newcommand{\ket}[1]{\mbox{$|#1\rangle$}}
\newcommand{\bra}[1]{\mbox{$\langle #1|$}}
\newcommand{\mypsfig}[2]{\psfig{file=#1,#2}}


\begin{abstract}

We present an efficient scheme which couples any designated pair of spins in
heteronuclear spin systems.
The scheme is based on the existence of Hadamard matrices.
For a system of $n$ spins with pairwise coupling, the scheme concatenates $cn$
intervals of system evolution and uses at most $c n^2$ pulses where 
$c \approx 1$.
Our results demonstrate that, in many systems, selective recoupling is
possible with linear overhead, contrary to common speculation that exponential
effort is always required.

\end{abstract}

\pacs{}

\section{Introduction}

In recent proposals to perform quantum computation in nuclear spin systems
using nuclear magnetic resonance (NMR)
techniques~\cite{Gershenfeld97,Cory97x,Cory97b,Chuang97e}, coupled logic
operations are performed using spin-spin couplings that occur naturally in
molecular systems.
While this is straightforward for small systems with a few spins,
generalization to complex molecular structures has been challenging.  
The complication is caused by the many spin-spin couplings which occur along
with the desired one.
This fundamental task to turn off spurious evolution is so difficult that,
coercing a complex system to {\em do nothing}~\cite{Jozsa98} -- ceasing all
evolution -- can be just as difficult as making it do something
computationally useful.

The task of turning off all couplings is known in the art of NMR as {\em
decoupling}; doing this for all but a select subset of couplings is known as
{\em selective recoupling}.  
A common method to perform these tasks is to interrupt the free evolution by
carefully chosen pulses.  These pulses are single spin-1/2 (qubit) operations
that transform the hamiltonian in the time between pulses in such a manner
that unwanted evolutions in consecutive time intervals cancel out each other.

Pulse sequences which perform selective recoupling are generally difficult to
find for a large system.  Each pulse simultaneously affects many coupling
terms in the hamiltonian.  To turn off all but one of the coupling terms,
these pulses have to satisfy many simultaneous requirements.
Ingenious sequences have been found for usual NMR
applications~\cite{Ernst94,Slichter,Mansfield} but they do not address the
problems relevant to quantum computation.  In usual NMR applications, the
structure of the spin systems is not known a-priori.  Therefore, pulse
sequences are designed to address all the spins together rather than
individual spins.
Quantum computation brings new requirements, and initial
efforts~\cite{Linden98} have been made to develop pulse sequences to satisfy
these needs; however, to-date, schemes have necessitated resources (such as
total number of pulses applied) exponential in the number of spins being
controlled.

In this paper, we present an {\em efficient} scheme to perform selective
recoupling.
In contrast to the situation with traditional NMR, this scheme addresses the
problem relevant to NMR quantum computation, in which the molecular structure
is assumed to be well-known, and spins are individually addressible.
The method is related to a class of well-known matrices called {\em Hadamard
matrices}.  
We derive from any $n \times n$ Hadamard matrix a pulse scheme that decouples
$n$ spins using $n$ time intervals and ${\cal O}(n^2)$ pulses.
This decoupling scheme can easily be modified  
(i) to remove Zeeman evolution and 
(ii) to implement selective recoupling.  
This completes the construction for $n$ spins whenever $n \times n$ Hadamard
matrices exist. 
When $n \times n$ Hadamard matrices do not exist, a scheme for $n$ spins can
still be constructed using larger existing Hadamard matrices.  
In doing so, an extra amount of effort is required, but this can be bounded
using existence properties of Hadamard matrices.
Altogether, the scheme requires $cn$ time intervals and less than 
$cn^2$ pulses, where $c \approx 1$ with upper bound $c \leq 2$.
Our method applies whenever the spins couple pairwise and have very
different Zeeman frequencies, such as in heteronuclear spin systems.

The paper is structured as follows.  
In Section~\ref{sec:statement}, we review relevant concepts in NMR quantum
computing and re-state the problem precisely. 
In Section~\ref{sec:scheme}, we first motivate the construction of the
decoupling scheme with examples, and then describe the general construction
related to Hadamard matrices.  Important properties of Hadamard matrices 
are summarized.  Modifications of the decoupling scheme to perform selective
recoupling are described.
We conclude with some general remarks and discussions of various properties 
and limitations of the scheme.  

\section{NMR quantum computing and the statement of the problem} 
\label{sec:statement}

In this section, we describe the NMR system and describe how a
universal set of (non-fault tolerant)
operations~\cite{Div95,Barenco95,Deutsch95}, namely, the single qubit
operations and the controlled-NOT gate, can be realized using basic NMR
primitives.

We shall consider a physical system which consists of a solution of identical
molecules.  Each molecule has $n$ non-magnetically equivalent nuclear spins
which serve as qubits.
A static magnetic field is applied externally along the $+\hat{z}$ direction.
This magnetic field splits the energy levels of the spin states aligned with
and against it.  This is described in the hamiltonian by the Zeeman terms,
which, in the energy eigenbasis, are given by
\be
	{\cal H}_{\rm Z} 
	= -\frac{1}{2} \sum_i \hbar \omega_i \sigma_z^{(i)} 
\,,
\label{eq:ham:zeeman}
\ee
where $i$ is the spin index, $\omega_i/2\pi$ is the {\em Zeeman
frequency} for the $i$-th spin, and $\sigma_z^{(i)}$ is the Pauli
matrix operating on the $i$-th spin.  
The convention $\hbar = 1$ is used for the rest of the paper.
The spins have very different Zeeman frequencies, a situation loosely termed 
as ``heteronuclear'' in this paper.

Nuclear spins can interact via the dipolar coupling~\cite{Slichter,Abragam}.
This is given by the hamiltonian
\be
	{\cal H}_{\rm d} = \sum_{i<j} g_{ij}^{\rm d} \lbL 
		\vec{\sigma}^{(i)} \cdot \vec{\sigma}^{(j)} 
		- 3 (\hat{r}_{ij} \cdot \vec{\sigma}^{(i)})  
		\otimes (\hat{r}_{ij} \cdot \vec{\sigma}^{(j)}) \rbL
\label{eq:dipolarfull}
\,,
\ee
where $\hat{r}_{ij}$ denotes the unit displacement vector from the $i$-th to
the $j$-th spin, and $g_{ij}^{\rm d}$ denotes the coupling constant between
them.
Spin-spin coupling can also be mediated by coupling to electrons.  This
indirect coupling has a tensor part and a scalar part.  The tensor part is
usually of the same form as ${\cal H}_{\rm d}$.  The scalar part is given by
the hamiltonian 
\be 
	{\cal H}_{\rm s} = \sum_{i<j} g_{ij}^{\rm s} 
			\vec{\sigma}^{(i)} \cdot \vec{\sigma}^{(j)} 
\label{eq:scalar}
\,.  
\ee 
If the molecules tumble fast and isotropically, dipolar coupling and
indirect tensor coupling will be averaged away; otherwise, the physics 
can be more complicated.  
However, in the presence of a strong external magnetic field, only the {\em
secular} part (terms that commute with ${\cal H}_{\rm Z}$) is
important~\cite{Slichter,Abragam}.
For a heteronuclear system, the resulting coupling becomes 
\be 
	{\cal H}_{\rm c} = \sum_{i<j} g_{ij}^{\rm c} \sigma_z^{(i)} \otimes
	\sigma_z^{(j)}
\label{eq:dipolar}
\,, 
\ee 
independent of the original form of coupling.  

Single qubit operations are performed by applying {\em pulsed} radio frequency
(RF) magnetic fields along some directions $\hat{\eta}$ perpendicular to the
static field.
To address the $i$-th spin, the frequency of the RF field is chosen to
approximate $\omega_i/2\pi$.
When the $\omega_i$'s are very different, very short pulses can be used, so
that during the pulses, all other evolutions are negligible except for the
rotation operator $e^{-i \frac{\theta}{2} \vec{\sigma}^{(i)} \cdot
\hat{\eta}}$ where $\theta$ is proportional to the pulse duration and the
power.  The Lie group of all single qubit operations can be generated by
rotations about $\hat{x}$ and $\hat{y}$.
Our scheme uses only rotations of $\theta = \pi$ along $\hat{x}$, which
implement $\sigma_x$ (up to an irrelevant overall phase) on the spins being
addressed.  We denote this operation by $X$, superscripted by the spin index
whenever appropriate.

Coupled operations such as controlled-phase-shift or controlled-NOT acting on
the $i$-th and the $j$-th spins can be performed given the primitive, 
\be 
      	Z\!\!Z_{ij} = 
	e^{-i \frac{\pi}{4} \sigma_z^{(i)} \otimes \sigma_z^{(j)}} 
\label{eq:zzij}
\,.
\ee 
For instance, a controlled-NOT from the $i$-th to the $j$-th spin
can be implemented by
\be
	e^{-i \frac{\pi}{4} \sigma_y^{(i)}} 
	e^{i \frac{\pi}{4} \sigma_x^{(i)}} 
	e^{i \frac{\pi}{4} \sigma_y^{(i)}} 
	e^{-i \frac{\pi}{4} \sigma_x^{(j)}} 
	e^{i \frac{\pi}{4} \sigma_y^{(j)}}
	Z\!\!Z_{ij} 
	e^{-i \frac{\pi}{4} \sigma_y^{(j)}}  
\,.
\ee

The ultimate goal is to be able to efficiently realize arbitrary quantum
operations on an $n$ spin system with arbitrary couplings.  In this paper, we
consider a more limited objective, which can now be stated precisely, using
the definitions of Eq.(\ref{eq:ham:zeeman}), Eq.(\ref{eq:dipolar}), and
Eq.(\ref{eq:zzij}):
\begin{quote} 
Given a heteronuclear system of $n$ spins with free evolution $e^{-i ({\cal
H}_{\rm Z} + {\cal H}_{\rm c}) t}$, controlled using typical RF pulses, how
can $Z\!\!Z_{ij}$ be implemented efficiently?
\end{quote}

\section{Construction of the scheme} 
\label{sec:scheme} 

We will first construct a decoupling scheme to remove the entire coupling term
${\cal H}_{\rm c}$ in the total evolution.  The scheme is derived from
Hadamard matrices, which will be reviewed.
Methods to remove ${\cal H}_{\rm Z}$ and to implement selective recoupling
will be described afterwards.

\subsection{Construction of the decoupling scheme} 
\label{sec:examples} 

To construct the decoupling scheme, we only consider ${\cal H}_{\rm c}$ in
the evolution.  Effects of ${\cal H}_{\rm Z}$ can be included later since all
matrix exponents commute.

To motivate the general construction, we analyze the simplest example of
decoupling two spins.  The evolution operator for an arbitrary duration $t$ is
given by $\tau = e^{-i g_{12}^{\rm c} t \sigma_z^{(1)} \otimes
\sigma_z^{(2)}}$.  The sequence of events, $\tau X^{(2)} \tau X^{(2)}$, known
as {\em refocusing} in NMR, will first couple and then decouple the spins.
This can be described mathematically as:
\bea
	&   & \tau (X^{(2)} \tau  X^{(2)}) 
\label{eq:refocus}
\\
	& = & e^{-i \theta \sigma_z^{(1)} \otimes \sigma_z^{(2)}} 
	    (\sigma_x^{(2)} 
	      e^{-i \theta \sigma_z^{(1)} \otimes \sigma_z^{(2)}} 
	     \sigma_x^{(2)})  
\\
	& = &  e^{-i \theta \sigma_z^{(1)} \otimes \sigma_z^{(2)}} 
		e^{-i \theta \sigma_z^{(1)} \otimes \sigma_x^{(2)} 
		\sigma_z^{(2)} \sigma_x^{(2)}}
\label{eq:third}
\\
	& = &  e^{-i \theta \sigma_z^{(1)} \otimes \sigma_z^{(2)}} 
	e^{-i \theta \sigma_z^{(1)} \otimes (-\sigma_z^{(2)})}
\label{eq:fourth}
\\
	& = &  e^{-i \theta \sigma_z^{(1)} \otimes \sigma_z^{(2)}} 
		e^{i \theta \sigma_z^{(1)} \otimes \sigma_z^{(2)}} 
\\
	& = &  I 
\eea
where $\theta = g_{12}^{\rm c} t$.  Eq.(\ref{eq:third}) is obtained using
Taylor series expansion of the matrix exponents and using $\sigma_x^2 = I$,
and Eq.(\ref{eq:fourth}) is obtained using anticommutivity of $\sigma_x$ and
$\sigma_z$.

The essential features of this decoupling procedure are:
\\ \noindent {\bf 1.} The pair of $X^{(2)}$ negates the sign of
$\sigma_z^{(2)}$ in the evolution between the pulses.
\\ \noindent {\bf 2.} The $X$ pulses make the signs of the $\sigma_z$
matrices of the two spins disagree for exactly half of the time.
\\ \noindent {\bf 3.} Since the coupling is bilinear in the $\sigma_z$
matrices, it is unchanged (negated) when the signs of the $\sigma_z$
matrices agree (disagree).  The $X$ pulses therefore negate the coupling for
exactly half of the time.
\\ \noindent {\bf 4.} Since the matrix exponents commute, negating the
coupling for exactly half of the time suffices for the evolution to be 
cancelled out. 

The most crucial point leading to decoupling is that, the signs of the
$\sigma_z$ matrices of the coupled spins, controlled by pairs of $X$ pulses,
disagree for half of the time.

In general, we consider pulse sequences which concatenate equal-time intervals
and use $X$ pulses to control the signs of the $\sigma_z$ of each spin.
The essential information on the signs can be represented by 
a ``sign matrix'' defined as follows.
The ``sign matrix'' of a pulse scheme for $n$-spins with $m$ time intervals is
the $n \times m$ matrix with the $(i,a)$ entry being the {\em sign} of
$\sigma_z^{(i)}$ in the $a$-th time interval.
These sign matrices have one-to-one correspondence with our restricted class
of pulse sequences.
We denote any sign matrix for $n$ spins by $S_n$. 
For example, the sequence in Eq.(\ref{eq:refocus}) can be represented by 
the sign matrix
\be
	S_2 = \left[ \begin{array}{cc}
	{+}&{+}
\\	{+}&{-} 
	\end{array} \right] 
\label{eq:s2}
\,. 
\ee 
The general construction of decoupling scheme is now reduced to finding 
sign matrices such that every two rows disagree in half of the entries.  

\begin{figure}[ht]
\begin{center}
\mbox{\psfig{file=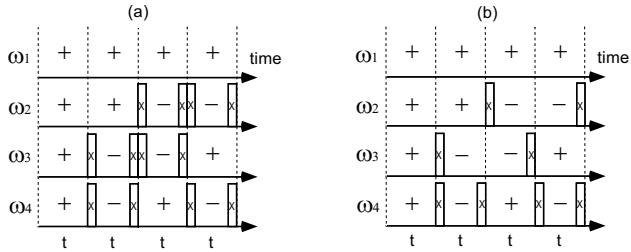,width=3.3in}}
\vspace*{2ex}
\caption{(a) Pulse sequence corresponding to Eq.(\ref{eq:dec4}).  From $S_4$,
each ``$-$'' sign in the $i$-th row and $a$-th column translates to two $X$
pulses at $\omega_i$ before and after the $a$-th time interval.
(b) Pulse sequence obtained from simplifying (a).  This corresponds to
Eq.(\ref{eq:dec4short}), and can be constructed directly from $S_4$ by
translating each change of sign in the $i$-th row to an $X$ pulse at
$\omega_i$.  A ``$-$'' sign at the end of the row also gives rise to an $X$
pulse at end of the last time interval.}
\label{fig:pulseseq}
\end{center}
\end{figure}

As a second example, we construct a decoupling scheme for four spins.  We
first find a correct sign matrix following the previous observations, and
then derive the corresponding pulse sequence.  For example, a possible 
sign matrix is given by 
\be 
	S_4 = \left[ \begin{array}{cccc}
	{+}&{+}&{+}&{+}
\\	{+}&{+}&{-}&{-}
\\	{+}&{-}&{-}&{+}
\\	{+}&{-}&{+}&{-}
	\end{array} \right] 
\label{eq:s4}
\,, 
\ee 
in which any two rows disagree in exactly two entries.  
$S_4$ can be converted to a pulse scheme by converting each column to a time
interval before and after which $X$ pulses are applied to spins (rows) given
by $-$'s.  No pulses are applied to spins (rows) with $+$'s.  The resulting
sequence,
\bea
	\tau 
	(X^{(3)} X^{(4)} \tau X^{(3)} X^{(4)})  
	(X^{(2)} X^{(3)} \tau X^{(2)} X^{(3)})  \times 
\nonumber
\\
	\hfill (X^{(2)} X^{(4)} \tau X^{(2)} X^{(4)})  
\label{eq:dec4}
\,, 
\eea
is the identity by construction and this can also be verified directly.  
Note that ${\cal H}_{\rm c}$ in $\tau = e^{-i {\cal H}_{\rm c} t}$ now denotes
the sum of six possible coupling terms for four spins.
Note also Eq.(\ref{eq:dec4}) is written in such a way that it corresponds
visually to the sign matrix, though the evolutions are actually in reverse
time order relative to $S_4$.  However, such ordering is irrelevant for
commuting evolutions.  Since $X^{(i)} X^{(i)} = I$, Eq.(\ref{eq:dec4}) can be
simplified to
\be
	\tau 
	(X^{(3)} X^{(4)} \tau X^{(4)})  
	(X^{(2)} \tau X^{(3)})  
	(X^{(4)} \tau X^{(2)} X^{(4)})  
\label{eq:dec4short}
\,.
\ee
This simplified pulse sequence can also be obtained directly from
Eq.(\ref{eq:s4}) by converting columns to time intervals and inserting
$X^{(i)}$ between intervals whenever the $i$-th row changes sign or 
whenever a $-$ sign reaches either end of the row.  Pulse sequences for
Eq.(\ref{eq:dec4}) and Eq.(\ref{eq:dec4short}) are shown in
Fig.~\ref{fig:pulseseq}.

The above scheme can be generalized to decouple $n$ spins with $m$ time
intervals as follows:
\begin{quote} 
Construct the $n \times m$ sign matrix $S_n$, with entries $+$ or $-$, such
that {\em any} two rows disagree in exacly half of the entries.  For
each $-$ sign 
in the $i$-th row and the $a$-th column, apply $X^{(i)}$ before and after the
$a$-th time interval.
\end{quote}
Because of the pulses, the sign of the $\sigma_z$ matrix for each spin in
each time interval is as given by the sign matrix.  The $\sigma_z$ matrices
of any two spins therefore have opposite signs for half of the time, so
that their coupling is negated for exactly half of the time, and the 
evolution is always cancelled.

For $n$ spins, $n \times m$ sign matrices which correspond to decoupling
schemes do not necessarily exist for arbitrary $m$, but they always exist for
large and special values of $m$.  A possible structure is:
\bea
\nonumber
	S_n = \left[ \begin{array}{cccccccccccc}
	{+}&{\cdots}&{+}&{+}&{\cdots}&{+}&{+}&{\cdots}&{+}&{+}&{\cdots}&{+}
\\	{+}&{\cdots}&{+}&{+}&{\cdots}&{+}&{-}&{\cdots}&{-}&{-}&{\cdots}&{+}
\\	{ }&{\cdots}&{ }&{ }&{\cdots}&{ }&{ }&{\cdots}&{ }&{ }&{\cdots}&{ }
\\	{+}&{\cdots}&{+}&{-}&{\cdots}&{-}&{+}&{\cdots}&{+}&{-}&{\cdots}&{-}
\\	{ }&{\cdots}&{ }&{ }&{\cdots}&{ }&{ }&{\cdots}&{ }&{ }&{\cdots}&{ } 
\\	{+}&{\cdots}&{-}&{+}&{\cdots}&{-}&{+}&{\cdots}&{-}&{+}&{\cdots}&{-}
	\end{array} \right] 
\,, 
\eea 
in which intervals are bifurcated when rows (spins) are added.  Such
bifurcation takes place whenever it is impossible to add an extra row that is
orthogonal to all the existing ones (``depletion'').  If such depletion occurs
frequently, the sign matrix will have exponential number of columns, and
decoupling will take exponential number of steps as $n$ increases.
The challenge is to find correct sign matrices with subexponential number of
columns. 
This difficult problem turns out to have solutions given by the Hadamard 
matrices, which will be decribed next.  

\subsection{Hadamard matrices}
\label{sec:Hadamard}

Hadamard matrices have applications in many areas such as the construction of
designs, error correcting codes and Hadamard
transformations~\cite{CRC96,Sloanehp,vanLint92,MacWilliams77}.

A Hadamard matrix of order $n$, denoted by $H(n)$, is an $n \times n$ matrix
with entries $\pm 1$, such that
\be
	H(n)H(n)^{T} = n I
\label{eq:ortho}
\,.
\ee
The rows are pairwise orthogonal, therefore any two rows agree in exactly half
of the entries.  Likewise columns are pariwise orthogonal.  We abbreviate
``$\pm 1$'' as ``$\pm$''.  $S_2$ and $S_4$ in Eqs.(\ref{eq:s2}) and
(\ref{eq:s4}) are simple examples of $H(2)$ and $H(4)$.  An example of $H(12)$
is given by
\bea
\nonumber
	H(12) = \left[ \begin{array}{cccccccccccc}
	{+}&{+}&{+}&{+}&{+}&{+}& {-}&{+}&{+}&{+}&{+}&{+}
\\	{+}&{+}&{+}&{-}&{-}&{+}& {+}&{-}&{+}&{-}&{-}&{+}
\\	{+}&{+}&{+}&{+}&{-}&{-}& {+}&{+}&{-}&{+}&{-}&{-}
\\	{+}&{-}&{+}&{+}&{+}&{-}& {+}&{-}&{+}&{-}&{+}&{-}
\\	{+}&{-}&{-}&{+}&{+}&{+}& {+}&{-}&{-}&{+}&{-}&{+} 
\\	{+}&{+}&{-}&{-}&{+}&{+}& {+}&{+}&{-}&{-}&{+}&{-}
\\	{-}&{+}&{+}&{+}&{+}&{+}& {-}&{-}&{-}&{-}&{-}&{-}
\\	{+}&{-}&{+}&{-}&{-}&{+}& {-}&{-}&{-}&{+}&{+}&{-}
\\	{+}&{+}&{-}&{+}&{-}&{-}& {-}&{-}&{-}&{-}&{+}&{+}
\\	{+}&{-}&{+}&{-}&{+}&{-}& {-}&{+}&{-}&{-}&{-}&{+}
\\	{+}&{-}&{-}&{+}&{-}&{+}& {-}&{+}&{+}&{-}&{-}&{-} 
\\	{+}&{+}&{-}&{-}&{+}&{-}& {-}&{-}&{+}&{+}&{-}&{-}
	\end{array} \right] 
\,.
\eea 

The following is a list of useful facts about Hadamard matrices (details and 
proofs omitted):

\begin{enumerate} 
\item {\em Equivalence}~~Permutations or negations of rows or columns of
Hadamard matrices leave the orthogonality condition invariant.  Two Hadamard
matrices are equivalent if one can be transformed to the other by a series of
such operations.  Each Hadamard matrix is equivalent to a {\em normalized}
one, which has only $+$'s in the first row and column.  For instance, $H(12)$
shown previously can be {\em normalized} by negating the 7-${\rm th}$ row
and column.

\item {\em Necessary conditions}~~$H(n)$ exists only for $n = 1$, $n = 2$ or
$n \equiv 0 \bmod 4$.  This is obvious if the matrix is normalized, and the
columns are permuted so that the first three rows become: 
\bea
\nonumber
	\left[ \begin{array}{cccccccccccc}
	{+}&{\cdots}&{+}&{+}&{\cdots}&{+}& {+}&{\cdots}&{+}&{+}&{\cdots}&{+}
\\	{+}&{\cdots}&{+}&{+}&{\cdots}&{+}& {-}&{\cdots}&{-}&{-}&{\cdots}&{-}
\\	{+}&{\cdots}&{+}&{-}&{\cdots}&{-}& {+}&{\cdots}&{+}&{-}&{\cdots}&{-}
\\	{\cdot}&{\cdot}&{\cdot}&{\cdot}&   {\cdot}&{\cdot}&{\cdot}&{\cdot}
	&{\cdot}&{\cdot}&{\cdot}&{\cdot}
	\end{array} \right] 
\,.
\eea 

\item {\em Hadamard's conjecture}~\cite{Hadamard93}~~$H(n)$ exists for every
$n \equiv 0 \bmod 4$.  This famous conjecture is verified for all $n < 428$.  

\item {\em Sylvester's construction}~\cite{Sylvester67}~~If $H(n)$ and $H(m)$ 
exist, then $H(nm)$ can be constructed as $H(n) \otimes H(m)$. 
In particular, $H(2^r)$ can be constructed as $H(2)^{\otimes r}$, which is
proportional to the matrix representation of the Hadamard transformation
for $r$ qubits.

\item {\em Paley's construction}~\cite{Paley33}~~Let $q$ be an odd prime
power.  If $q \equiv 3 \bmod 4$, then $H(q+1)$ exists; if $q \equiv 1 \bmod 4$,
then $H(2(q+1))$ exists.  

\item {\em Numerical facts}~\cite{CRC96}~~
For an arbitrary integer $n$, let $\underline{n}$ and $\overline{n}$ be the
largest and smallest integers that satisfy $\underline{n} < n \leq
\overline{n}$ with {\em known} $H(\overline{n})$ and $H(\underline{n})$.  We
define the ``gap'', $\delta_n$, to be $\overline{n} - \underline{n}$ (see
Fig.~\ref{fig:gap}).
For $n\leq 1000$, $H(n)$ is known for every possible order except for 6 cases,
and the maximum gap is 8.  For $n \leq 10000$, $H(n)$ is unknown for 192
possible orders and the maximum gap is 32.

\begin{figure}[ht]
\begin{center}
\mbox{\psfig{file=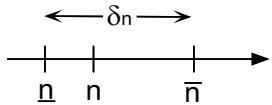,width=1.4in}}
\vspace*{2ex}
\caption{The gap $\delta_n$ between two existing orders of Hadamard matrices.}
\label{fig:gap}
\end{center}
\end{figure}

\end{enumerate}

The importance of the full connection to Hadamard matrices will become 
clear after we construct the scheme for an arbitrary number of spins 
in the next section.  

\subsection{General Construction of Scheme} 

In this section, a decoupling scheme for an arbitrary number of spins is
constructed.  Variations of the scheme to remove Zeeman evolution and to
implement selective recoupling are also constructed.  The requirements of the
scheme are discussed.

Recall that $\overline{n}$ is the minimum integer no smaller than $n$ with
known $H(\overline{n})$, and $m$ is the number of columns in a sign matrix.
For any variations of the scheme for $n$ spins, $m_n$ is defined to be the
minimum number of columns in any valid sign matrix and it represents the
number of time intervals needed for the intended operation.

To construct a decoupling scheme for $n$ spins when $H(n)$ does not
necessarily exist, any $H(\overline{n})$ with $\overline{n}-n$ rows omitted
can be used as the sign matrix since subsets of rows of $H(\overline{n})$ are
still pairwise orthogonal.
In other words, any $n \times \overline{n}$ submatrix of $H(\overline{n})$ is
a valid sign matrix for decoupling $n$ spins, and $m_n = \overline{n}$.

To remove both ${\cal H}_{\rm Z}$ and ${\cal H}_{\rm c}$, we use the fact that
the Zeeman term for the $i$-th spin is linear in $\sigma_z^{(i)}$, and
negating $\sigma_z^{(i)}$ for half of the time results in no net Zeeman
evolution for the $i$-th spin.  Therefore, Zeeman evolution for all spins can
be removed if the sign matrix has identically zero row sum.
Such a sign matrix can be constructed by starting with a {\em normalized}
$H(\overline{n})$ and excluding the first row of $H(\overline{n})$ in the sign
matrix.  Since a normalized $H(\overline{n})$ has only $+$'s in the first row,
all other rows have zero row sums by orthogonality.
Such construction is possible unless $n = \overline{n}$, in which case 
construction should start with $H(\overline{n+1})$.  Therefore, $m_n =
\overline{n}$ if $n < \overline{n}$ and $m_n = \overline{n+1}$ otherwise.

To implement selective recoupling between the $i$-th and the $j$-th spins, the
sign matrix should have equal $i$-th and $j$-th rows but any other two rows
should be orthogonal.
The coupling term $g_{ij}^{\rm c} \sigma_z^{(i)} \otimes \sigma_{\rm z}^{(j)}$
never changes sign and that coupling is implemented, while all other couplings
are removed.
The sign matrix can be obtained from a normalized $H(\overline{n})$ by
replacing the $1$-st row with the $j$-th row, and replacing the $j$-th row
with the $i$-th row.  This scheme also removes ${\cal H}_{\rm Z}$ and $m_n =
\overline{n}$.
To implement $Z\!\!Z_{ij}$, the duration of each interval $t$ is chosen
to satisfy $g_{ij}^{\rm c} \overline{n} t = \pi/4$.  Note that the total time
used to implement $Z\!\!Z_{ij}$ is the shortest possible, since the coupling
is always ``on''.

\begin{figure}[ht]
\begin{center}
\mbox{\psfig{file=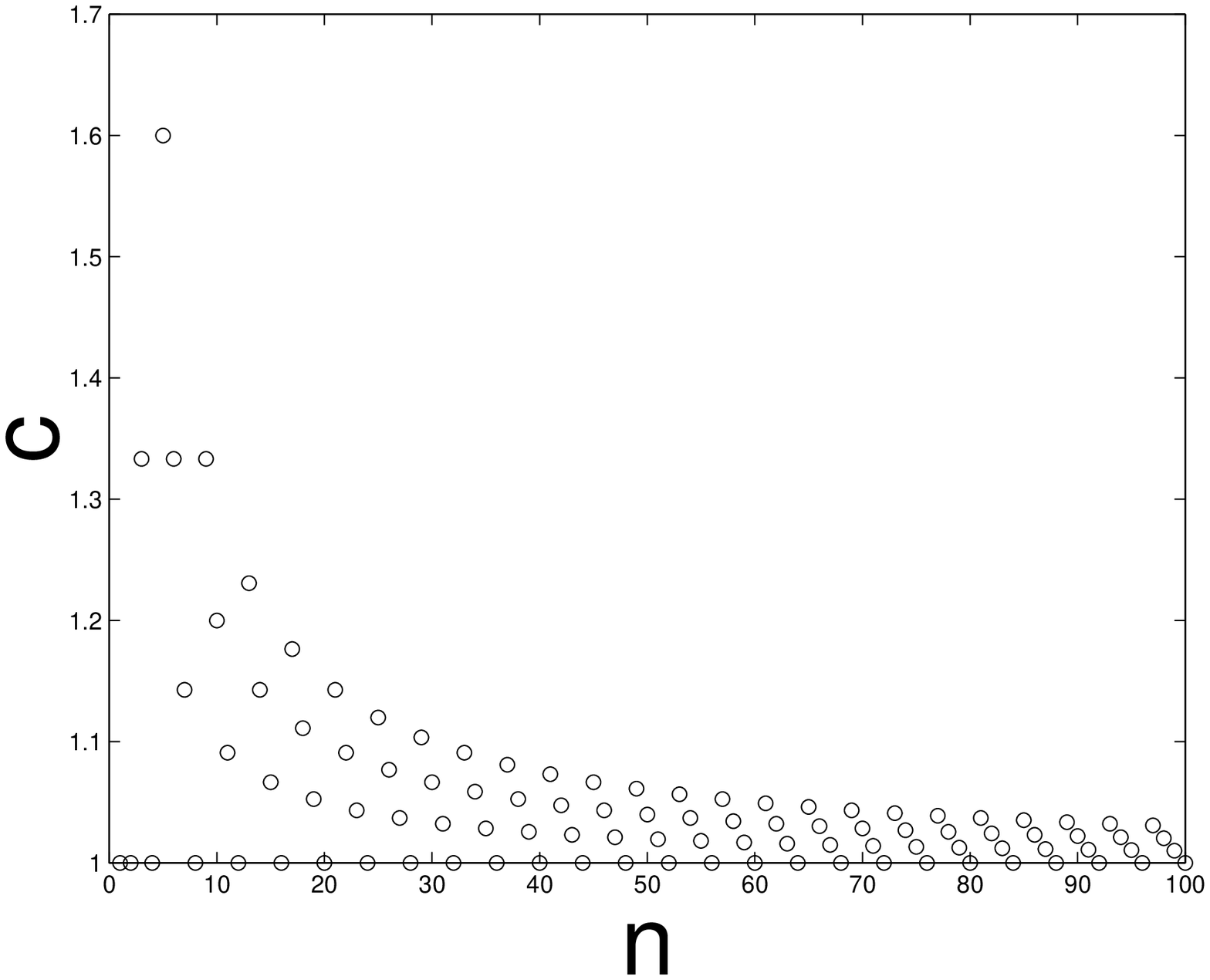,width=1.7in}\psfig{file=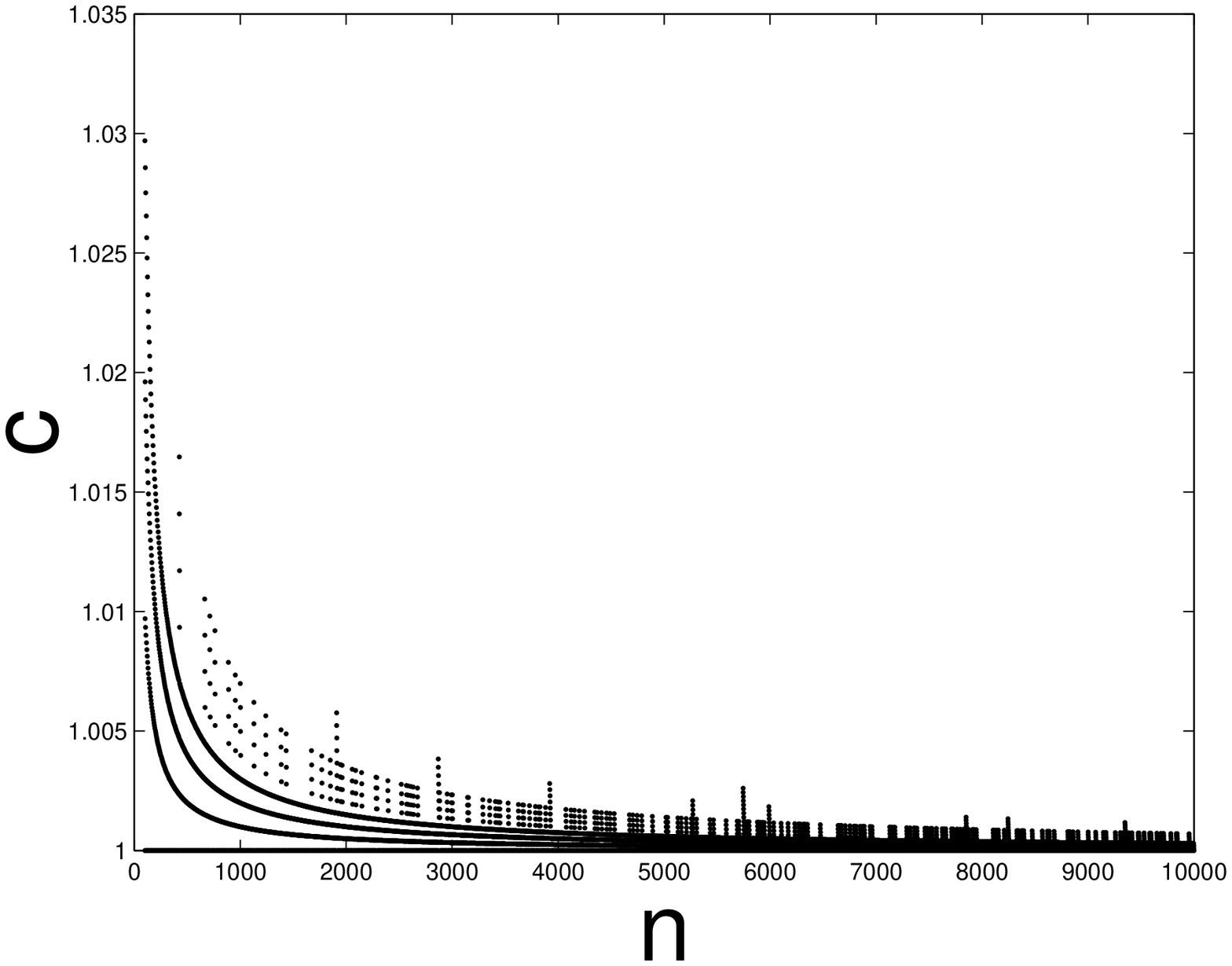,width=1.74in}}
\vspace*{2ex}
\caption{Plots of $c$ vs $n$, where $cn = \overline{n} = m_n$ is the minimun
number of time intervals required to perform decoupling or selective
recoupling for an $n$-spin system.  $c$ for $n \leq 100$ and $101 \geq n \leq
10000$ are plotted separately.}
\label{fig:c}
\end{center}
\end{figure}

The scheme requires $m_n$ time intervals.  Consequently, it requires at
most $n m_n$ pulses, since $XX=I$ and the $X$ pulses are only used in pairs.  
The remaining question is: how does $m_n$ depend on $n$?
For simplicity, we consider $\overline{n}$ in place of $m_n$. 
If Hadamard matrices exist and can be constructed for all orders,
$\overline{n} = n$.
However, some Hadamard matrices are missing, either because no construction
methods are known or they simply cannot exist.
Due to missing Hadamard matrices, $\overline{n} = cn$ where $c \geq 1$.  
We argue in the following that the scheme is still very efficient.  
First of all, we prove $c < 2$.  For each $n$, there exists $r$ such that 
$2^{r-1} \!  \leq n < 2^r$.  Since $H(2^r)$ exists by Sylvester's
construction, $cn = \overline{n} \leq 2^r \!\! < 2n$.
We now show that $c$ is close to the ideal value 1 in most cases, due to the
existence of Hadamard matrices of orders other than powers of 2.
This is why the full connection to Hadamard matrices is important.  
In Fig.~\ref{fig:c}, $c$ as a function of $n$ is plotted for $n \leq 10000$.  
Within this technologically relevant range of $n$, $c$ deviates significantly
from 1 only for a few exceptional values of $n$ when $n$ is small.
One arrives at the same conclusion by considering the gap $\delta_n =
\overline{n} - \underline{n} > \overline{n} - n$, which bounds the extra
number of intervals needed in the scheme caused by missing Hadamard matrices.
For $n \leq 10000$, $\delta_n/n \ll 1$ except for the few exceptional values
of $n$ as a numerical fact.
For completeness, we present arguments for $c \approx 1$ for {\em arbitrarily
large} $n$ in Appendix~\ref{sec:largen}.  
This is based on Paley's construction and the prime number theorem. 
Finally, if Hadamard's conjecture is proven, $\delta_n \leq 3$ $\forall n$.

\section{Conclusion} 

We reduce the problem of decoupling and selective recoupling in heteronuclear
spin systems to finding sign matrices which is further reduced to finding
Hadamard matrices.
While the most difficult task of constructing Hadamard matrices is not
discussed in this paper, solutions already exist in the literature. 
Even more important is that the connection to Hadamard matrices results in
very efficient schemes.

Some properties of the scheme are as follows.  
First of all, the scheme is optimal in the following sense.  
The rows of Hadamard matrices and their negations form the codewords of first
order Reed-Muller codes, which are {\em perfect
codes}~\cite{vanLint92,MacWilliams77}.
It follows that, for each Hadamard matrix, it is impossible to add an extra
row which is orthogonal to all the existing ones.
Therefore, for a given $n$, $m_n = \overline{n}$ is in fact the minimum number
of time intervals necessary for decoupling or recoupling, if one restricts 
to the class of pulse sequences considered.
Second, the scheme applies for arbitrary duration of the time intervals.  
This is a consequence of the commutivity of all the terms in the hamiltonian,
which in turn comes from the large separations of the Zeeman frequencies
compared to the coupling constants.  
Spin systems can be chosen to satisfy this condition.
Finally, disjoint pairs of spins can couple in parallel. 

We outline possible simplifications of the scheme for systems with restricted
range of coupling.
For example, a linear spin system with $n$ spins but only $k$-nearest
neighbor coupling can be decoupled by a scheme for $k$ spins only.  
The $i$-th row of the $n \times \overline{k}$ sign matrix can be chosen to be
the $r$-th row of $H(\overline{k})$, where $i \equiv r \bmod k$.
Selective recoupling can be implemented using a decoupling scheme for $k+1$
spins.  The sign matrix is constructed as in decoupling using
$H(\overline{k+1})$ but the rows for the spins to be coupled are chosen to be
the $k+1$-th row different from all existing rows~\cite{kplus1}.
This method involving periodic boundary conditions generalizes to other
spatial structures.  The size of the scheme depends on $k$ and the exact 
spatial structure but not on $n$.

The scheme has several limitations.  
First of all, it only applies to systems in which spins can be individually
addressed by short pulses and coupling has the simplified form given by
Eq.(\ref{eq:dipolar}).
These conditions are essential to the simplicity of the scheme.  They can all
be satisfied if the Zeeman frequencies have large separations.
Second, generalizations to include couplings of higher order than bilinear
remain to be developed.
Furthermore, in practice, RF pulses are inexact and have finite
durations, leading to imperfect transformations and residual errors.

The present discussion is only one example of a more general issue, that the
naturally occuring hamiltonian in a system does not directly give rise to
convenient quantum logic gates or other computations such as simulation of
quantum systems~\cite{Terhal98}.
Efficient conversion of the given system hamiltonian to a useful form is
necessary and is an important challenge for future research.

\section{Acknowledgments}

This work was supported by DARPA under contract DAAG55-97-1-0341 and Nippon
Telegraph and Telephone Corporation (NTT).  D.L. acknowledges support of an
IBM Co-operative Fellowship.  We thank Hoi-Fung Chau, Kai-Man Tsang, Hoi-Kwong
Lo, Alex Pines, Xinlan Zhou and Lieven Vandersypen for helpful comments.

 

\begin{thebibliography}{10}
\bibitem{kplus1} With a periodic sign matrix, setting the 
$j$-th row to be the $i$-th row also recoupling the $j$-th and the
$i+k$-th spins when $j>i$.  Similar problem occurs when $j<i$.   
%
Therefore a row different from all existing ones is used.  
%
This is not necessary in linear systems using a more complicated
scheme, but it is necessary for other structures such as a planar system.
\end{thebibliography}

\appendix

\section{Upper bounds for $\overline{n}$}
\label{sec:largen}

An argument for $c \approx 1$ for large $n$ is presented using Paley's
construction (mentioned in Section~\ref{sec:Hadamard}), known results on
primes in intervals and the prime number theorem for arithmetic progressions.

Let $\pi(x)$ be the number of primes which satisfy $2 \leq p \leq x$.  
For $x<67$, $x/(\log x - 1/2) < \pi(x) < x/(\log x - 3/2)$~\cite{Rosser62}. 
It follows that there exists a prime between $n$ and $n(1+\epsilon)$ 
for $\epsilon > 2/\log n$.  
Applying Paley's construction, $H(p+1)$ or $H(2(p+1))$ exists depending
on whether $p \equiv 3 \bmod 4$ or $p \equiv 1 \bmod 4$.  
Therefore, $\overline{n} \leq n(1+\epsilon)+1$ or $\overline{n} \leq
2(n(1+\epsilon)+1)$ respectively.

The worse of the upper bounds $\overline{n} \leq 2(n(1+\epsilon)+1)$ resulting
from $p \equiv 1 \bmod 4$ can be improved.  Note that there are at least $r$
primes between $n$ and $n(1+\epsilon)^r$.  If the primes that equal $3 \bmod
4$ and $1 \bmod 4$ are randomly and uniformly distributed, the probability to
find a prime which equals $3 \bmod 4$ between $n$ and $n(1+\epsilon)^r$ is
larger than $1-2^{-r}$.
This assumption is true due to the prime number theorem for arithmetic 
progressions~\cite{Davenport67}.  
Let $\pi(x,a,q)$ denotes the number of primes in the arithmetic progression
$\{a, a+q, a+2q, \ldots\}$ which satisfy $2 \leq p \leq x$.  It is known that
$\pi(x,3,4) \approx \pi(x,1,4)$.  Therefore, with probability larger than
$1-2^{-r}$, $\overline{n} \leq n(1+\epsilon)^r+1$, implying $c \leq
(1+\epsilon)^r+1/n \approx 1$ for large $n$.

\end{document}